\documentclass[conference]{IEEEtran}
\IEEEoverridecommandlockouts

\usepackage[utf8]{inputenc}
\usepackage[T1]{fontenc}
\usepackage{amsmath,amsfonts,amssymb}
\usepackage{graphicx}
\usepackage{booktabs}
\usepackage{url}
\usepackage{cite}
\usepackage{hyperref}
\usepackage{xcolor}
\usepackage{enumitem}
\usepackage{stfloats}
\usepackage{array, booktabs, tabularx} 
\usepackage{adjustbox} 
\usepackage{multirow}
\usepackage{makecell}
\usepackage{orcidlink}
\usepackage{hyperref}
\hypersetup{
    colorlinks=true,
    linkcolor=blue,
    filecolor=magenta,      
    urlcolor=blue,
    citecolor=blue
}

\makeatletter
\def\footnoterule{%
  \kern-5pt
  \hrule \@width .5\columnwidth \@height .4pt
  \kern 4.6pt
}
\makeatother

\usepackage{times}
\usepackage{fancyhdr}

\fancypagestyle{IEEEtitlepagestyle}{%
  \fancyhf{}
  \fancyhead[L]{\footnotesize Preprint}
  
}

\title{A Brain Wave Encodes a Thousand Tokens:\\
Modeling Inter-Cortical Neural Interactions\\ 
for Effective EEG-based Emotion Recognition
}

\author{%
\IEEEauthorblockN{%
Nilay~Kumar\orcidlink{0009-0001-3104-6779}\textsuperscript{\dag},
Priyansh~Bhandari\orcidlink{0009-0006-6515-8046},
G.~Maragatham\orcidlink{0000-0003-1589-0571}}
\IEEEauthorblockA{Department of Computational Intelligence, \\
SRM Institute of Science and Technology, KTR\\
\texttt{\{nl9459, pr6479, maragatg\}@srmist.edu.in}}
\thanks{\textsuperscript{\dag}Corresponding author: Nilay Kumar (nl9459@srmist.edu.in).}
}

\begin{document}
\maketitle

% Abstract
\begin{abstract}
Human emotions are difficult to convey through words and are often abstracted in the process; however, electroencephalogram (EEG) signals can offer a more direct lens into emotional brain activity. Recent studies show that deep learning models can process these signals to perform emotion recognition with high accuracy. However, many existing approaches overlook the dynamic interplay between distinct brain regions, which can be crucial to understanding how emotions unfold and evolve over time, potentially aiding in more accurate emotion recognition. To address this, we propose RBTransformer, a Transformer-based neural network architecture that models inter-cortical neural dynamics of the brain in latent space to better capture structured neural interactions for effective EEG-based emotion recognition. First, the EEG signals are converted into Band Differential Entropy (BDE) tokens, which are then passed through Electrode Identity embeddings to retain spatial provenance. These tokens are processed through successive inter-cortical multi-head attention blocks that construct an electrode × electrode attention matrix, allowing the model to learn the inter-cortical neural dependencies. The resulting features are then passed through a classification head to obtain the final prediction. We conducted extensive experiments, specifically under subject-dependent settings, on the SEED, DEAP, and DREAMER datasets, over all three dimensions, Valence, Arousal, and Dominance (for DEAP and DREAMER), under both binary and multi-class classification settings. The results demonstrate that the proposed RBTransformer outperforms all previous state-of-the-art methods across all three datasets, over all three dimensions under both classification settings. The source code is available at: \href{https://github.com/nnilayy/RBTransformer}{https://github.com/nnilayy/RBTransformer}.
\end{abstract}

% Keywords

% Introduction
\section{Introduction}
Emotions are complex psychophysiological responses to internal or external stimuli and are deeply embedded in human cognition, shaping how individuals feel, form mental states, and react emotionally to their surroundings \cite{james1884emotion,schachter1962cognitive,izard1984emotion}. These responses can be expressed through either non-physiological modalities, such as facial expressions, gestures, speech, and language \cite{ekman1971constants,poria2017review,dmello2015review}, which are often subject to voluntary control and may mask what a person is truly experiencing, or through physiological modalities, including heart rate, skin conductance, and especially neural activity \cite{calvo2010affect,koelstra2012deap}, which provide a more direct and involuntary window into how people respond to stimuli and how emotions evolve internally over time.

\begin{figure}
    \centering
    \includegraphics[width=\columnwidth]{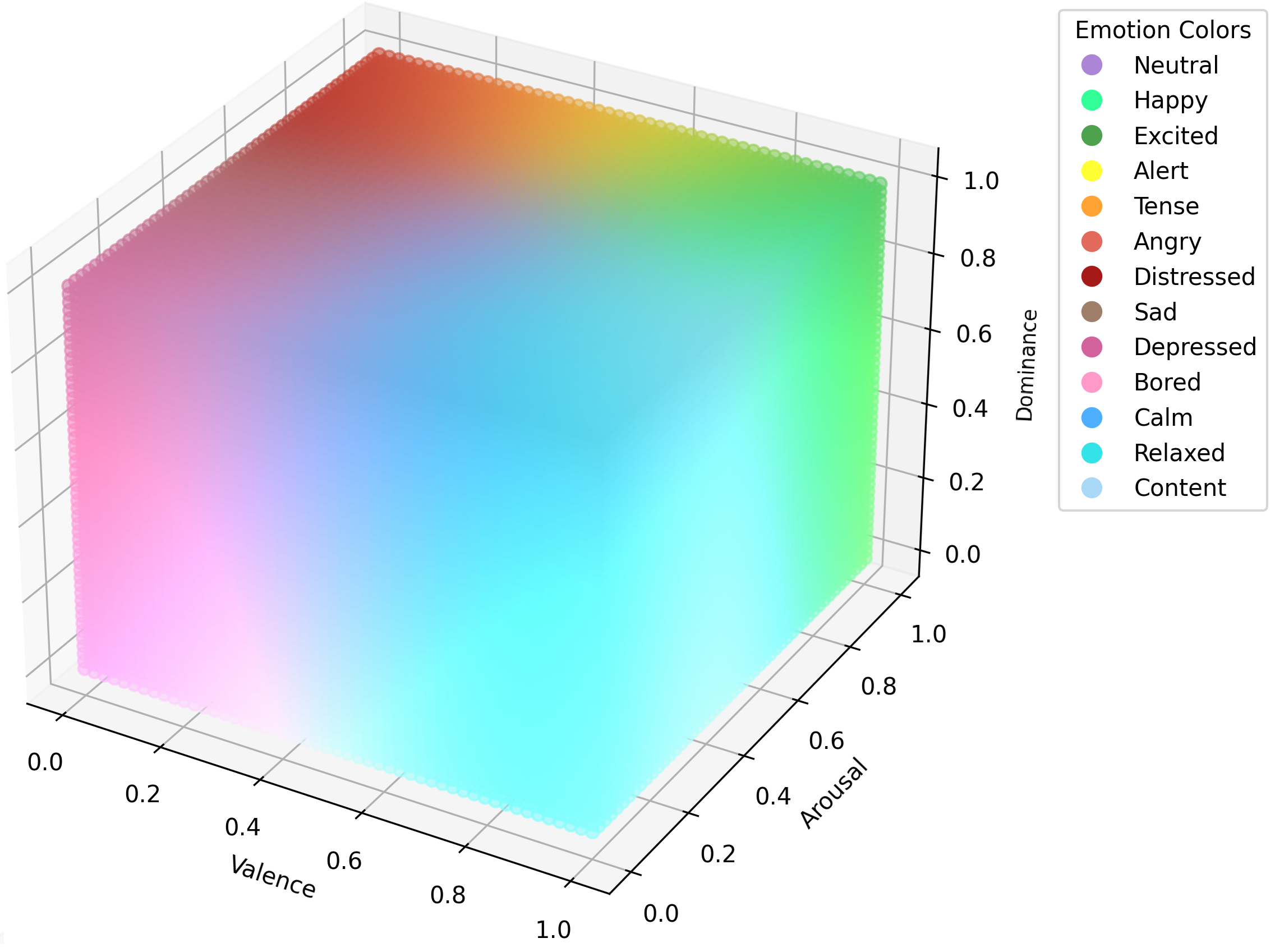}
    \caption{Visual representation of a wide range of emotions in the three-dimensional Valence–Arousal–Dominance (VAD) space.}
    \label{fig:combined_emotions}
\end{figure}

\begin{figure}
    \centering
    \includegraphics[width=\columnwidth]{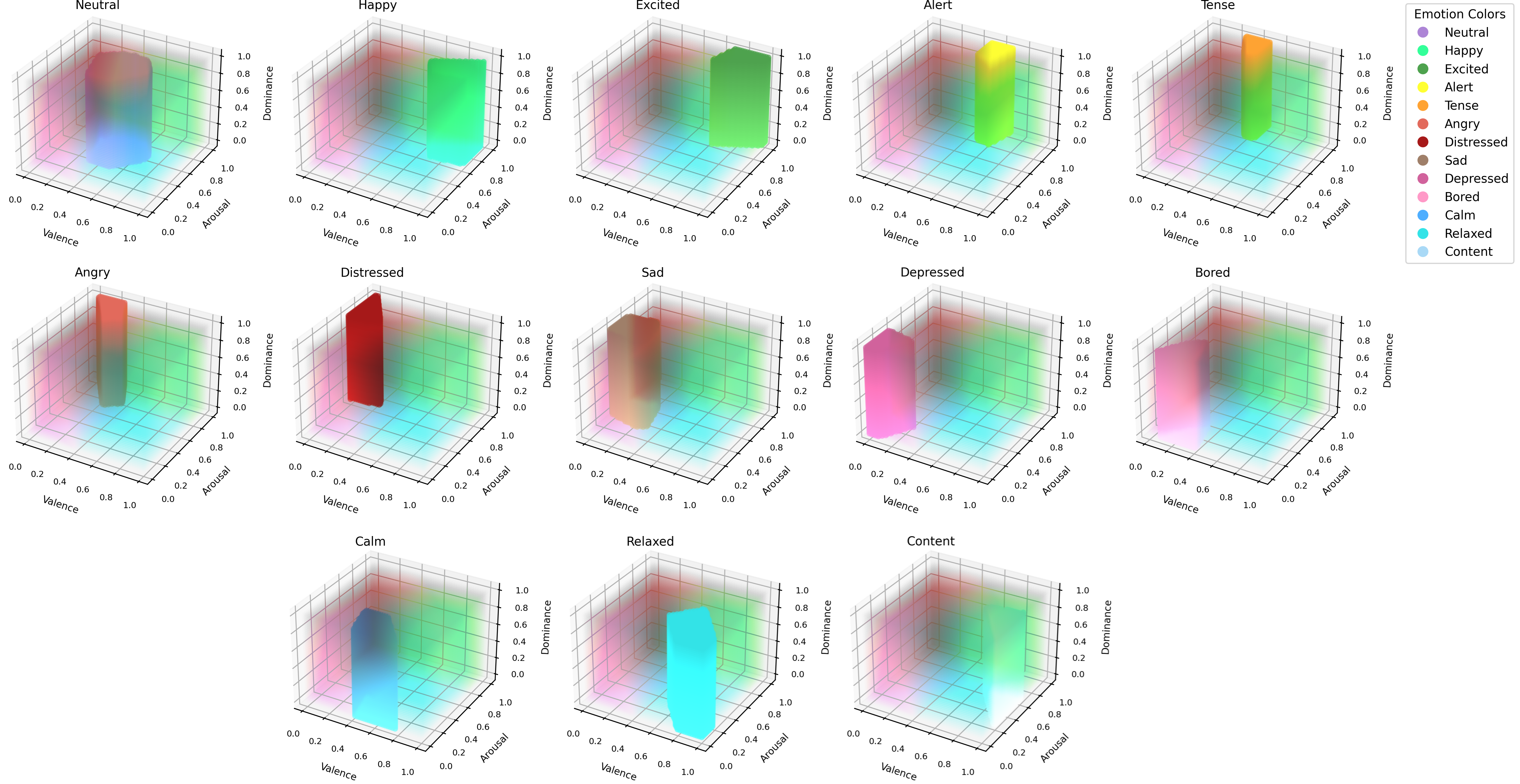}
    \caption{Individual emotion representations within the three-dimensional Valence–Arousal–Dominance (VAD) space, showcasing the unique characteristics of each emotion.}
    \label{fig:individual_emotions}
\end{figure}

Quantifying and recognizing emotions plays a crucial role in healthcare and clinical contexts, helping in the understanding of emotional dysregulation and impaired affective expression, which are important for the diagnosis and treatment of medically diagnosed conditions, as defined by the DSM-5~\cite{dsm5} and ICD-11 classification systems~\cite{icd11}, including neuropsychiatric conditions (e.g., depression, anxiety) and broader neurodevelopmental and psychotic disorders (e.g., autism spectrum disorder, schizophrenia)~\cite{kragel2016emotions,adolphs2017emotion}. To quantify and enable such modelling, emotions can be modelled using either discrete or dimensional models. Discrete models can classify emotions into only a fixed set of labels (e.g., happiness, sadness, anger, fear, disgust, surprise)~\cite{ekman1992argument}. Dimensional models are more general and can describe any emotion by expressing them along three linear scales, Valence, Arousal, and Dominance (VAD)~\cite{russell1980circumplex,mehrabian1996pleasure}, which aligns more closely with how emotions are naturally experienced and support representations that generalize across individuals and contexts~\cite{barrett1999structure}. As a result, any emotion represented using a discrete model can be plotted within the VAD dimensional model, shown in Figures~\ref{fig:combined_emotions} and~\ref{fig:individual_emotions}. Valence represents the degree of pleasantness, Arousal represents the level of alertness or activation, and Dominance represents the sense of control over emotion. To capture raw and involuntary neural activity, various brain–computer interface (BCI) techniques are used like functional magnetic resonance imaging (fMRI), magnetoencephalography (MEG), and near-infrared spectroscopy (NIRS)~\cite{lindquist2012brain,doi2013nirs}. However, electroencephalography (EEG) is still the most widely used technique because of its non-invasive nature, cost-effectiveness, and high temporal resolution, making it well-suited for real-time emotion monitoring. 

% Place both images one below the other, one-column wide

Traditionally, emotion recognition using electroencephalographic (EEG) signals has relied on manually extracting features through domain-specific preprocessing and feature extraction techniques, subsequently passing them to shallow machine learning models. The extracted features are generally derived from one of three primary domain representations: (i) time-domain (e.g., Hjorth parameters~\cite{hjorth1970eeg}, Higher Order Crossing (HOC)~\cite{petrantonakis2009emotion}), (ii) frequency-domain (e.g., Differential Entropy (DE)~\cite{shi2013de}, Power Spectral Density (PSD)~\cite{redwan2024psd}), and (iii) time–frequency domain (e.g., Discrete Wavelet Transform (DWT)~\cite{kumar2017dwt}, Continuous Wavelet Transform (CWT)~\cite{ende1998cwt}). Using these representations, features were manually handcrafted and fed into shallow classifiers. For example, Li et al. manually extracted 18 time-domain features (9 Hjorth parameters and 9 nonlinear features) from gamma-band EEG signals after applying Common Spatial Pattern (CSP) filtering and used a linear Support Vector Machine (SVM) for emotion recognition~\cite{li2009emotion}. Patil et al. applied Empirical Mode Decomposition (EMD) to denoise EEG signals and extracted features using HOC for emotion recognition~\cite{patil2015emd}. Shi et al. extracted features across five frequency bands using DE and passed them to SVM for final classification~\cite{shi2013de}, and similarly, Duan et al. extracted features using DE from multichannel EEG data and passed them into a hybrid k-Nearest Neighbors (k-NN) and SVM classifier~\cite{duan2013differential}.

However, over the past decade, deep neural networks (DNNs) have steadily outperformed traditional machine learning models across a range of fields, including computer vision~\cite{lecun1989backpropagation}, natural language processing~\cite{vaswani2017attention}, and human-computer interaction~\cite{koelstra2012deap}. A similar progression was observed in the domain of EEG-based emotion recognition, when Zheng and Lu introduced Deep Belief Network (DBN)~\cite{zheng2015investigating}, one of the first deep learning models for affective emotion recognition. The DBN learned hierarchical representations from DE features and outperformed previous state-of-the-art machine learning approaches. In the following years, subsequent studies, such as those by Huang and Zhao~\cite{huang2021bidcnn} and Rudakov~\cite{rudakov2021multi}, made use of Convolutional Neural Networks (CNNs)~\cite{lecun1989backpropagation} to capture spatial dependencies across EEG electrodes. By projecting signals into structured 2D topographies or frequency-based brain maps, CNNs enabled more effective modeling of spatial information. In parallel with the development of CNN-based approaches, the sequential and autoregressive nature of EEG signals led studies such as those by Zhang et al.~\cite{zhang2020spatial} and Ma et al.~\cite{ma2019emotion} to explore Recurrent Neural Networks (RNNs)~\cite{elman1990finding} and Long Short-Term Memory (LSTM)~\cite{hochreiter1997long} architectures, as these were better suited for capturing temporal dependencies and tracking emotional state transitions over time. Naturally, to capture both the spatial and temporal richness of EEG signals, studies such as those by Shen et al.~\cite{shen2020eeg4d} and Yang et al.~\cite{yang2018parallelcrnn} introduced hybrid architectures combining CNNs and RNNs (or LSTMs). During the development of such models, studies such as those by Song et al.~\cite{song2018dgcnn} and Yin et al.~\cite{yin2021fusion} made use of Graph Neural Networks (GNNs)~\cite{scarselli2009gnn} to more accurately ground the spatial structure of EEG data neurophysiologically, effectively modeling neural and inter-electrode dependencies as compared to the fixed-grid representations used in earlier approaches.

In recent years, studies have turned to Transformer and attention-based models~\cite{vaswani2017attention} for their ability to capture long-range temporal dependencies and leverage a non-recurrent architecture that enables global context access, parallel computation, and more efficient training. Studies such as those by Tao et al.~\cite{tao2020eeg} apply spatial and temporal attention to identify informative electrodes and time steps for emotion classification. Xiao et al.~\cite{xiao2020multi} extend this by adding frequency-band attention to weigh spectral features. Liu et al.~\cite{liu2022emotion} employ dual channel attention within a 3D CNN to emphasize key spatio-temporal patterns. While these and other EEG-based emotion recognition models leverage attention mechanisms to highlight salient features within individual domains, such as spatial (electrode), spectral (frequency band), or temporal (time slice), they fundamentally treat each electrode’s signal as an independent input stream, and attention in these architectures primarily functions as a filtering mechanism, adaptively weighting channels or segments based on their relative informativeness for the task. However, none of the existing studies or model architectures take into account inter-electrode interactions, which are more neurologically grounded in how emotions emerge and are processed across cortical regions, and which can lead to more accurate and robust affective state recognition.

To explore and address this issue, we introduce RBTransformer, a Transformer-based neural network architecture that models these inter-cortical neural interactions in the latent space for EEG-based emotion recognition. RBTransformer first converts raw EEG signals into Band Differential Entropy (BDE) tokens, which are then passed through an Electrode Identity Embedding layer, which allows the model to retain awareness of each electrode’s unique identity and ordering. These representations are then passed through a stack of Inter-Cortical Multi-Head Attention Blocks, which allow each electrode to directly interact with every other electrode using an electrode × electrode attention matrix. And finally, these features are passed through a classification head to get the final class prediction. This architecture mimics the recurrent exchange of information across cortical regions, allowing RBTransformer to capture both inter-cortical dependencies and localized temporal dynamics, without relying on handcrafted features or sequential modeling at all. In brief, the primary contributions of this paper are as follows:

\begin{enumerate}
    \item We introduce RBTransformer, a Transformer-based architecture that explicitly models inter-cortical neural interactions through a dedicated multi-head attention mechanism. By enabling structured communication between EEG electrodes and incorporating frequency-aware BDE tokens along with Electrode Identity Embeddings, the model captures both localized saliency and global inter-regional dependencies without relying on handcrafted features or explicit temporal modeling.

    \item We conduct extensive experiments and demonstrate that RBTransformer achieves state-of-the-art performance across all three benchmark datasets, SEED, DEAP, and DREAMER, across all dimensions (Valence, Arousal, and Dominance for DEAP and DREAMER), under both binary and multi-class emotion classification settings.

    \item We also present t-SNE visualizations and confusion matrices to demonstrate that the model effectively segregates emotional classes in the latent space and maintains consistent discriminative performance across classes.
        
\end{enumerate}

The rest of the paper is organized as follows: Section II reviews related work. Section III describes data preprocessing steps and model architecture for RBTransformer. Section IV outlines datasets, evaluation metrics, and training configurations. Section V presents the results of extensive experiments and highlights the effectiveness of the proposed model. Finally, Section VI concludes the paper.

% Methodology
\section{Methodology}
In this section, we present the end-to-end pipeline used to preprocess the raw EEG data for RBTransformer and provide a detailed explanation of RBTransformer's model architecture.

\subsection{Dataset Preprocessing}

\begin{figure}
    \centering
    \includegraphics[width=\linewidth]{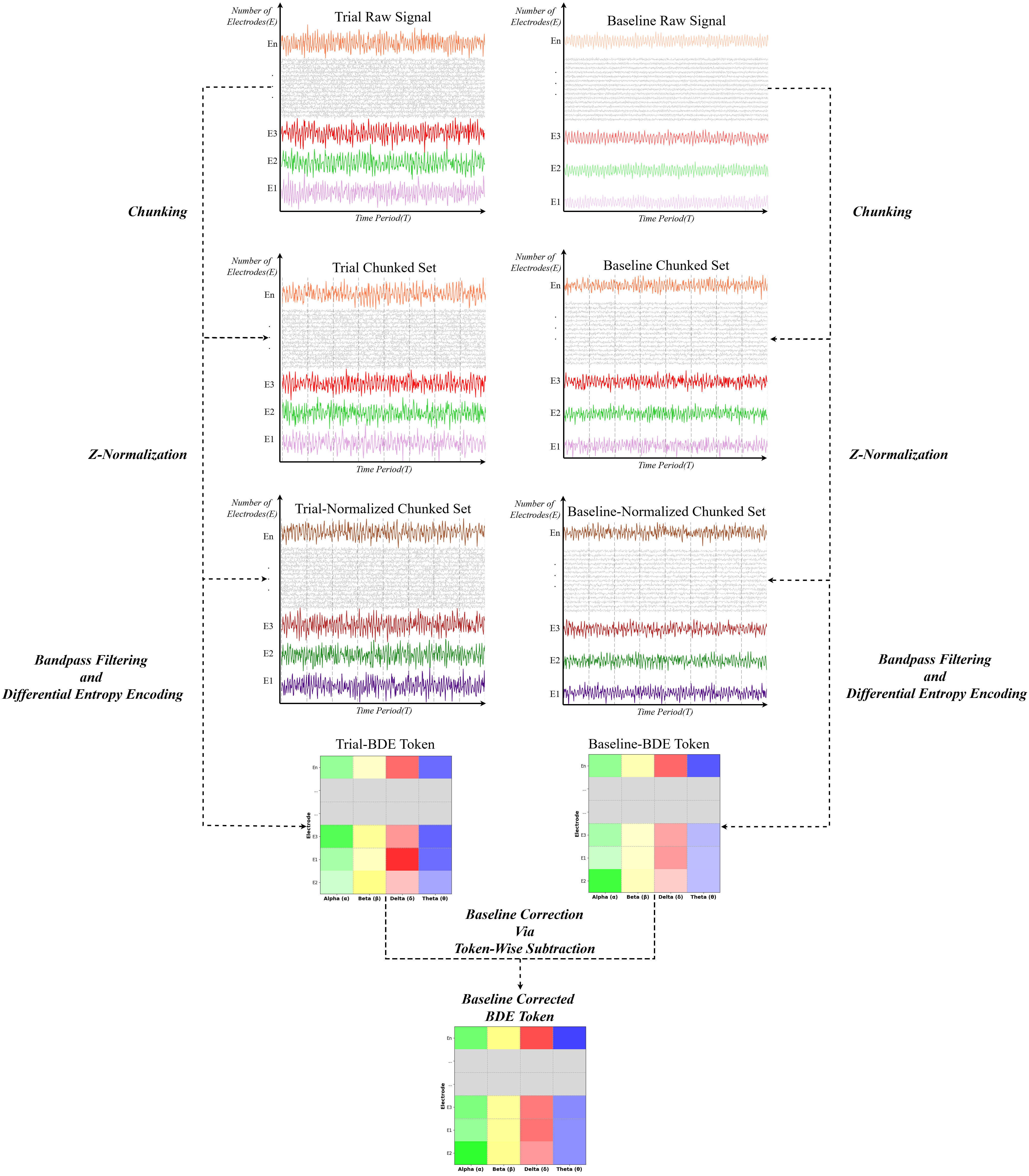}
    \caption{Preprocessing pipeline applied across EEG datasets for RBTransformer.}
    \label{fig:eeg_preprocessing}
\end{figure}

We use the SEED~\cite{zheng2015investigating}, DEAP~\cite{koelstra2012deap}, and DREAMER~\cite{katsigiannis2018dreamer} datasets for our experiments and apply a consistent preprocessing pipeline across all three datasets. The complete EEG preprocessing pipeline is presented in Fig.~\ref{fig:eeg_preprocessing}, and the entire workflow is explained as follows.

Each dataset \(\mathit{X}\) is made up of continuous multichannel EEG recordings, where each recording includes both a baseline segment \(\mathit{X}_B\) and a trial stimulus-induced response segment \(\mathit{X}_T\), as shown in Eq.~\eqref{eq:dataset_structure}.

\begin{equation}
\mathit{X} = [\mathit{X}_B, \mathit{X}_T]
\label{eq:dataset_structure}
\end{equation}

For each recording \(i\), the pair \((\mathit{X}_{Bi}, \mathit{X}_{Ti})\) corresponds to the baseline segment \(\mathit{X}_{Bi} \in \mathbb{R}^{\mathit{C} \times \mathit{L}_B}\) and the trial segment \(\mathit{X}_{Ti} \in \mathbb{R}^{\mathit{C} \times \mathit{L}_T}\), where \(\mathit{C}\) is the number of EEG channels, and \(\mathit{L}_B\), \(\mathit{L}_T\) are the respective time points in the baseline and trial segments.

Next, these baseline and trial signals are chunked into smaller windows. For the trial segments, a sliding window of size \(512\) with a stride of \(117\) is applied to each \(\mathit{X}_{Ti}\). Each \(\mathit{X}_{Ti}\) is chunked into \(\mathit{M}\) overlapping windows of size \(512\), denoted as \(\mathit{X}_{Ti} = \{ \mathit{X}_{Ti}^{(1)}, \mathit{X}_{Ti}^{(2)}, \dots, \mathit{X}_{Ti}^{(M)} \}\), where each \(\mathit{X}_{Ti}^{(j)} \in \mathbb{R}^{\mathit{C} \times 512}\). To chunk baseline segments, the average signal is computed per trial as shown in Eq.~\eqref{eq:avg_baseline}:
\begin{equation}
\bar{\mathit{X}}_{Bi} = \frac{1}{\mathit{M}} \sum_{j=1}^{\mathit{M}} \mathit{X}_{Bi}^{(j)}
\label{eq:avg_baseline}
\end{equation}

Here, each \(\mathit{X}_{Bi}\) is chunked into \(\mathit{M}\) non-overlapping windows of size \(128\), denoted as \(\mathit{X}_{Bi} = \{ \mathit{X}_{Bi}^{(1)}, \mathit{X}_{Bi}^{(2)}, \dots, \mathit{X}_{Bi}^{(M)} \}\), where each \(\mathit{X}_{Bi}^{(j)} \in \mathbb{R}^{\mathit{C} \times 128}\). This process yields a set of averaged and chunked baseline signals \(\bar{\mathit{X}}_B = \{ \bar{\mathit{X}}_{B1}, \bar{\mathit{X}}_{B2}, \dots, \bar{\mathit{X}}_{Bn} \}\).

To reduce inter-subject and inter-channel variability, z-score normalization is applied independently to both the trial chunks \( \mathit{X}_{Ti}^{(j)} \in \mathbb{R}^{\mathit{C} \times 512} \) and the averaged baseline segments \( \bar{\mathit{X}}_{Bi} \in \mathbb{R}^{\mathit{C} \times 128} \). Let \( \mathit{M} \in \mathbb{R}^{\mathit{C} \times L} \) denote either a trial chunk \( \mathit{X}_{Ti}^{(j)} \) or a baseline segment \( \bar{\mathit{X}}_{Bi} \), where \( L = 512 \) for trial chunks and \( L = 128 \) for baseline segments. The normalization is performed per channel \( c \in \{1, 2, \dots, \mathit{C}\} \), shown in equations Eqs.~\eqref{eq:mean}--\eqref{eq:norm}:

\begin{equation}
\mu_c = \frac{1}{L} \sum_{t=1}^{L} \mathit{M}[c, t]
\label{eq:mean}
\end{equation}

\begin{equation}
\sigma_c = \sqrt{ \frac{1}{L} \sum_{t=1}^{L} \left( \mathit{M}[c, t] - \mu_c \right)^2 }
\label{eq:std}
\end{equation}

\begin{equation}
\hat{\mathit{M}}[c, t] = \frac{ \mathit{M}[c, t] - \mu_c }{ \sigma_c }, \quad \forall t \in \{1, \dots, L\}
\label{eq:norm}
\end{equation}

Here, \(\mu_c\) and \(\sigma_c\) are the mean and standard deviation across the \(L\) time points in channel \(c\), and \(\hat{\mathit{M}} \in \mathbb{R}^{\mathit{C} \times L}\) denotes the normalized matrix with zero mean and unit variance per channel. This process yields the set of normalized trial chunks \(\hat{\mathit{X}}_T = \{ \hat{\mathit{X}}_{T1}^{(1)}, \hat{\mathit{X}}_{T1}^{(2)}, \dots, \hat{\mathit{X}}_{Tn}^{(M)} \}\) and the set of normalized baseline segments \(\hat{\bar{\mathit{X}}}_B = \{ \hat{\bar{\mathit{X}}}_{B1}, \hat{\bar{\mathit{X}}}_{B2}, \dots, \hat{\bar{\mathit{X}}}_{Bn} \}\), which serve as standardized inputs for the subsequent feature extraction stage.

Band Differential Entropy (BDE)~\cite{zheng2015investigating} is then calculated for each normalized trial chunk and the corresponding normalized baseline segment. Bandpass filters are used to extract sub-band signals corresponding to four EEG frequency bands: Theta (4--8~Hz), Alpha (8--13~Hz), Beta (13--30~Hz), and Gamma (30--45~Hz). The differential entropy for each frequency band \(b\) and channel \(c\) is computed using the following formulation, shown in Eq.~\eqref{eq:bde_general}:

\begin{equation}
\text{BDE}_c^{(b)} = \frac{1}{2} \log_2 \left( 2\pi e \cdot \text{Var} \left( \hat{\mathit{S}}^{(b)}[c] \right) \right)
\label{eq:bde_general}
\end{equation}

Here, \(\hat{\mathit{S}}^{(b)}[c]\) denotes the filtered signal in frequency band \(b\) for channel \(c\), which may correspond to either a trial segment \(\hat{\mathit{X}}_{Ti}^{(j, b)}[c]\) or a baseline segment \(\hat{\bar{\mathit{X}}}_{Bi}^{(b)}[c]\). \(\text{Var}(\cdot)\) represents the sample variance of the sub-band signal, capturing the frequency-specific energy distribution across EEG channels. This yields trial BDE feature tokens \(\mathit{F}_i^{(j)} \in \mathbb{R}^{\mathit{C} \times 4}\) and baseline BDE feature tokens \(\bar{\mathit{F}}_i \in \mathbb{R}^{\mathit{C} \times 4}\). The full sets are denoted as \(\mathcal{F} = \{ \mathit{F}_1^{(1)}, \mathit{F}_1^{(2)}, \dots, \mathit{F}_n^{(M)} \}\) and \(\bar{\mathcal{F}} = \{ \bar{\mathit{F}}_1, \bar{\mathit{F}}_2, \dots, \bar{\mathit{F}}_n \}\), respectively.

Finally, to eliminate subject-specific trends, remove static background activity, and isolate emotion-related neural activity, baseline correction is applied by subtracting the baseline BDE feature token \(\bar{F}_i \in \mathbb{R}^{C \times 4}\) from the corresponding trial BDE feature token \(F_i^{(j)} \in \mathbb{R}^{C \times 4},\) as shown in Eq.~\eqref{eq:bde_correction}:

\begin{equation}
\tilde{F}_i^{(j)} = F_i^{(j)} - \bar{F}_i
\label{eq:bde_correction}
\end{equation}

Here, $\tilde{F}_i^{(j)} \in \mathbb{R}^{C \times 4}$ denotes the final baseline-corrected BDE token for the $j^\text{th}$ chunk of the $i^\text{th}$ trial. Collectively, this yields the fully preprocessed dataset $\tilde{\mathcal{F}} = \{ \tilde{F}_1^{(1)}, \tilde{F}_1^{(2)}, \dots, \tilde{F}_n^{(M)} \}$, which serves as the input for training and inference for RBTransformer.

\subsection{Model Architecture}
\begin{figure*}
\centering
\includegraphics[width=\textwidth]{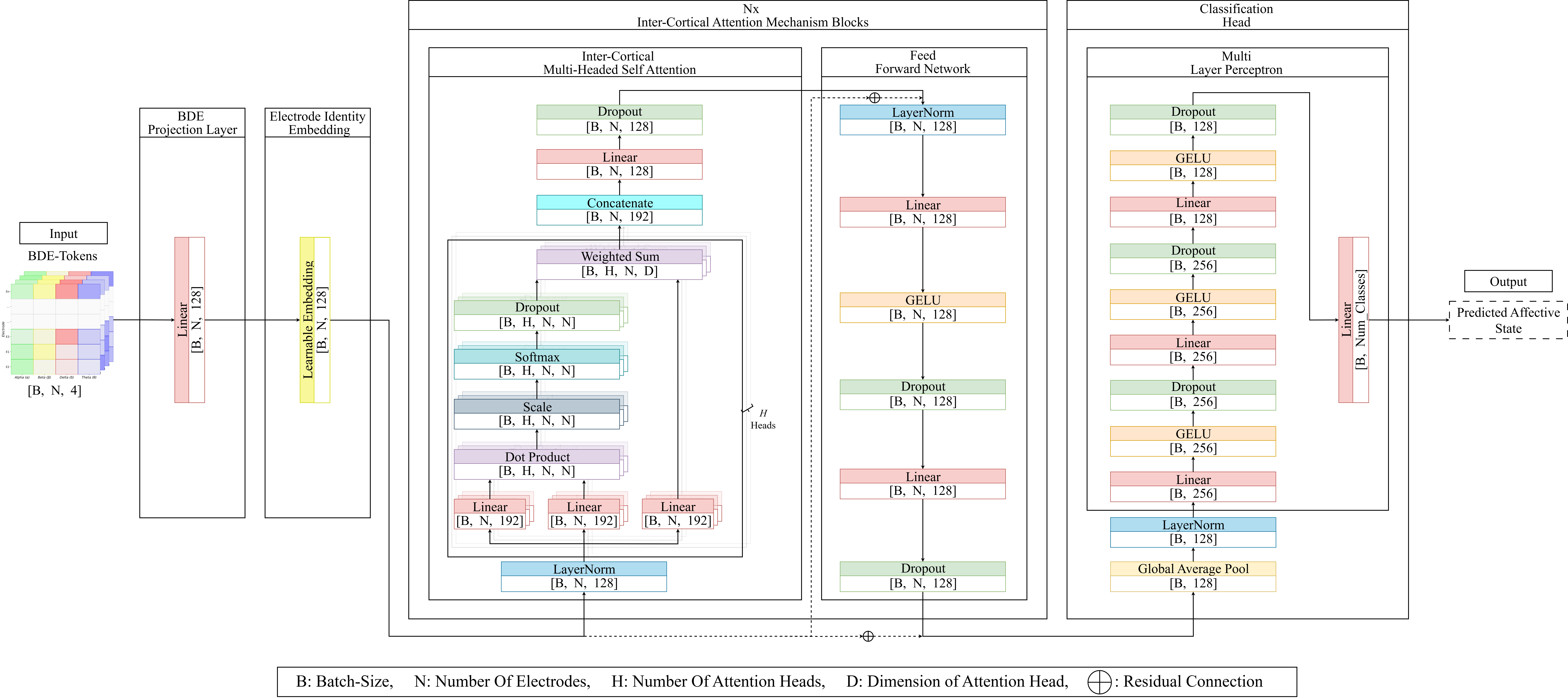}
\caption{Schematic architecture diagram of RBTransformer implementing inter-cortical attention.}
\label{fig:model-arch}
\end{figure*}

The RBTransformer model is an attention‐based architecture tailored for EEG‐based emotion recognition. It operates on the baseline‐normalized Band Differential Entropy (BDE) tokens, where each token encodes the deviation of each electrode’s spectral energy from its baseline across canonical EEG bands. These inputs are passed through the following sequence of processing components, depicted in Figure~\ref{fig:model-arch}, which illustrates the complete architecture layout from BDE token input to classification output.

\textbf{(1) BDE Feature Projection Layer:}  
A batch $B$ of preprocessed BDE tokens $\tilde{\mathit{F}} \in \mathbb{R}^{B \times C \times 4}$ is passed through a shared linear projection layer that maps each 4-dimensional BDE vector into a $d_{\mathit{model}}$-dimensional embedding vector for all electrodes $C$, as shown in Eq.~\eqref{eq:bde_projection}:

\begin{equation}
\mathit{Z} = \tilde{\mathit{F}}\, \mathit{W}_{\text{proj}} + \mathit{b}_{\text{proj}}
\label{eq:bde_projection}
\end{equation}

Here, a dot product is computed between each 4-dimensional BDE token and the shared projection matrix $\mathit{W}_{\text{proj}} \in \mathbb{R}^{4 \times d_{\mathit{model}}}$, followed by the addition of a learnable bias vector $\mathit{b}_{\text{proj}} \in \mathbb{R}^{d_{\mathit{model}}}$,  
which results in each BDE token being transformed into a richer representational space of dimension $d_{\mathit{model}}$.

\textbf{(2) Electrode Identity Embedding Layer:}  
To support inter-cortical modeling, the attention mechanism relies on a fixed electrode-to-electrode structure, which is achieved by assigning a unique learnable embedding to each electrode to retain its spatial information. Accordingly, after BDE projection, a learnable identity embedding $\mathit{E}_{\text{identity}} \in \mathbb{R}^{1 \times C \times d_{\mathit{model}}}$ is added to the projected tensor $\mathit{Z}$, broadcast across the batch, as shown in Eq.~\eqref{eq:identity_embedding}:

\begin{equation}
\mathit{Z}' = \mathit{Z} + \mathit{E}_{\text{identity}}
\label{eq:identity_embedding}
\end{equation}

which is then followed by a Dropout for regularization, shown in Eq.~\eqref{eq:dropout_layer}:

\begin{equation}
\tilde{\mathit{Z}} = \mathit{Dropout}(\mathit{Z}')
\label{eq:dropout_layer}
\end{equation}

\textbf{(3) Inter-Cortical Attention Mechanism:}  
Following the addition of electrode identity embeddings, the intermediate features then enter the model's core module, the Inter-Cortical Attention Mechanism. For this, RBTransformer uses a stack of attention blocks, each made up of a Multi-Head Self-Attention module and a Feedforward Network, both of which are explained in detail in the following sections. 

\quad\textbf{(I) Multi-Head Self-Attention (MHSA):}  
The Electrode Identity Embedded tokens are then passed through the Multi-Head Self-Attention (MHSA) module, which serves as the core component for modeling the inter-cortical neural interactions in the latent space, simulating how different regions of the brain exchange information. In this, input tokens are first passed through a normalization layer, which helps to stabilize the training by ensuring that the features are scaled consistently across the embedding dimension. This is shown in Eq.~\eqref{eq:mhsa_layernorm}.

\begin{equation}
\hat{\mathit{Z}} = \text{LayerNorm}(\tilde{\mathit{Z}})
\label{eq:mhsa_layernorm}
\end{equation}

The normalized tokens $\tilde{\mathit{Z}} \in \mathbb{R}^{B \times C \times d_{\mathit{model}}}$ are then linearly projected into three learnable representations, queries (Q), keys (K), and values (V). This operation is performed independently across $H$ attention heads, each of dimensionality $d_{\mathit{head}}$. Each head maintains its own set of QKV projections, $\mathit{Q}, \mathit{K}, \mathit{V} \in \mathbb{R}^{B \times H \times C \times d_{\mathit{head}}}$, allowing the model to capture multi-perspective inter-regional dependencies from the same neural signal simultaneously. This linear projection is computed via a dot product between the normalized input tokens and a learnable weight matrix $\mathit{W}_{\mathit{qkv}} \in \mathbb{R}^{d_{\mathit{model}} \times 3H d_{\mathit{head}}}$, as shown in Eq.~\eqref{eq:qkv_projection}.

\begin{equation}
[\mathit{Q}, \mathit{K}, \mathit{V}] = \hat{\mathit{Z}} \cdot \mathit{W}_{\mathit{qkv}}
\label{eq:qkv_projection}
\end{equation}

After this comes the inter-cortical attention mechanism, which computes an attention matrix of shape (electrodes $\times$ electrodes). This attention matrix $\mathit{Attention} \in \mathbb{R}^{B \times H \times C \times C}$ enables each electrode to attend to every other electrode, effectively capturing inter-regional neural interactions. It is calculated by taking the dot product between the Queries (Q) and Keys (K) matrices, scaled by the square root of the head dimension $\sqrt{d_{\mathit{head}}}$, as shown in Eq.~\eqref{eq:attention_scores}.

\begin{equation}
\label{eq:attention_scores}
\mathit{Attention} = \text{Softmax}\left(\frac{\mathit{Q} \mathit{K}^{\top}}{\sqrt{d_{\mathit{head}}}}\right)
\end{equation}

These attention weights are applied to the values~$(\mathit{V})$ to obtain the head-specific outputs. The outputs from all heads are concatenated and projected back to the original embedding dimension using a linear transformation $\mathit{W}_{\mathit{out}} \in \mathit{R}^{\mathit{H} d_{\mathit{head}} \times d_{\mathit{model}}}$, as shown in Eq.~\eqref{eq:mhsa_output}:

\begin{equation}
\label{eq:mhsa_output}
\mathit{Z}_{\mathit{attn}} = \text{Concat}(\mathit{Attention} \cdot \mathit{V}) \cdot \mathit{W}_{\mathit{out}}
\end{equation}

Finally, a residual connection is applied, along with dropout regularization, to produce the final output of the MHSA module $\mathit{Z}_{\mathit{MHSA}} \in \mathit{R}^{\mathit{B} \times \mathit{C} \times d_{\mathit{model}}}$, as given in Eq.~\eqref{eq:mhsa_residual}:
\begin{equation}
\label{eq:mhsa_residual}
\mathit{Z}_{\mathit{MHSA}} = \tilde{\mathit{Z}} + \text{Dropout}(\mathit{Z}_{\mathit{attn}})
\end{equation}

\quad\textbf{(II) Feedforward Network (FFN):}  
The output of the MHSA block $\mathit{Z}_{\mathit{MHSA}}$, is then passed through a Feedforward Network (FFN), which enhances the model’s ability to capture complex local patterns and non-linear relationships within each signal. Specifically, the FFN applies a stack of linear transformations and dropout layers to project the input representation and promote regularization, while a $\mathit{GELU}$ activation introduces non-linearity to better model complex local dependencies. This sequence refines the representation at each electrode. Finally, a residual connection adds the FFN output back to the original input $\mathit{Z}_{\mathit{MHSA}}$, helping preserve the original information flow and mitigate vanishing gradients during training, as shown in Eqs.~\eqref{eq:ffn_ffn}–\eqref{eq:ffn_residual}.

\begin{align}
\label{eq:ffn_ffn}
\mathit{Z}_{\mathit{FFN}} &= \mathit{Dropout} \big( 
    \mathit{W}_{\mathit{out}} \cdot \mathit{Dropout} \big( \\
    &\quad \mathit{GELU} \big( 
        \mathit{W}_{\mathit{in}} \cdot \mathit{Z}_{\mathit{MHSA}} + \mathit{b}_{\mathit{in}} 
    \big) \big) + \mathit{b}_{\mathit{out}} \big) \notag \\
\label{eq:ffn_residual}
\mathit{Z}_{\mathit{AttnBlock}} &= \mathit{Z}_{\mathit{MHSA}} + \mathit{Z}_{\mathit{FFN}}
\end{align}

Here, $\mathit{Z}_{\mathit{AttnBlock}} \in \mathbb{R}^{\mathit{B} \times \mathit{C} \times \mathit{embed\_dim}}$ represents the output of one attention block. $\mathit{W}_{\text{in}} \in \mathbb{R}^{\mathit{embed\_dim} \times \mathit{hidden\_dim}}$ and $\mathit{W}_{\text{out}} \in \mathbb{R}^{\mathit{hidden\_dim} \times \mathit{embed\_dim}}$ are learnable weight matrices, while $\mathit{b}_{\text{in}} \in \mathbb{R}^{\mathit{hidden\_dim}}$ and $\mathit{b}_{\text{out}} \in \mathbb{R}^{\mathit{embed\_dim}}$ are the corresponding learnable bias vectors.

\textbf{(4) Classification Head:}  
Finally, after the input is processed through multiple attention blocks, it enters the classification head, which converts the contextualized electrode representations into emotion predictions. The output tensor $\mathit{Z}_{\mathit{AttnBlockFinal}}$ is first aggregated using Global Average Pooling (GAP) across the electrode dimension. The resulting pooled vector, $\mathit{Z}_{\text{GAP}} \in \mathbb{R}^{B \times \mathit{embed\_dim}}$, is then normalized via LayerNorm (LN) to obtain $\mathit{Z}_{\text{LN}} \in \mathbb{R}^{B \times \mathit{embed\_dim}}$, which is subsequently passed through a linear layer to produce the final class logits, $\mathit{y}_{\text{logits}} \in \mathbb{R}^{B \times \mathit{num\_classes}}$, as shown in Eqs.~\eqref{eq:cls_gap}–\eqref{eq:cls_output}:

\begin{align}
\label{eq:cls_gap}
\mathit{Z}_{\text{GAP}} &= \frac{1}{C} \sum_{i=1}^{C} \mathit{Z}_{\mathit{AttnBlock}}[:, i, :] \\
\label{eq:cls_ln}
\mathit{Z}_{\text{LN}} &= \text{LayerNorm}(\mathit{Z}_{\text{GAP}}) \\
\label{eq:cls_output}
\mathit{y}_{\text{logits}} &= \mathit{W}_{\text{logits}} \cdot \mathit{Z}_{\text{LN}} + \mathit{b}_{\text{logits}}
\end{align}

Here, $\mathit{W}_{\text{logits}} \in \mathbb{R}^{\mathit{embed\_dim} \times \mathit{num\_classes}}$ and $\mathit{b}_{\text{logits}} \in \mathbb{R}^{\mathit{num\_classes}}$ are the learnable weights and bias matrices of the final classification head.

% Experimental Setup
\section{Experimental Setup}
In this section, we describe the datasets used for training and benchmarking, the metrics used for performance evaluation, and the training configuration for RBTransformer.

\subsection{Datasets}
We perform experiments on three standardized EEG emotion recognition benchmarks, SEED~\cite{zheng2015investigating}, DREAMER~\cite{katsigiannis2018dreamer}, and DEAP~\cite{koelstra2012deap}. The details of each dataset are explained as follows.

\begin{figure*}[htbp]
    \centering
    \includegraphics[width=0.75\textwidth]{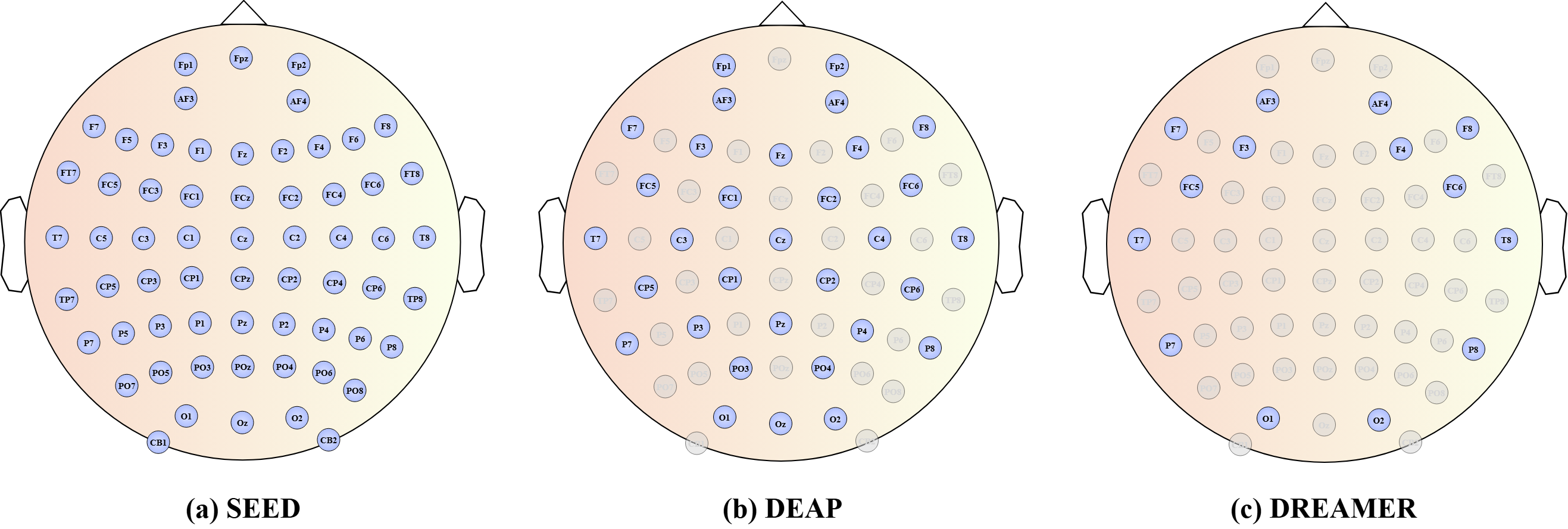}
    \caption{
    Electrode layouts for (a) SEED~\cite{zheng2015investigating} with 62 electrodes, (b) DEAP~\cite{koelstra2012deap} with 32 electrodes, and (c) DREAMER~\cite{katsigiannis2018dreamer} with 14 electrodes. Blue circles indicate active electrodes used in each dataset; Grey circles denote unused electrodes.
    }
    \label{fig:electrode_comparison}
\end{figure*}

\textbf{(1) SEED:} The SEED dataset~\cite{zheng2015investigating} contains EEG recordings from 15 subjects (7 male, 8 female) with an average age of 23.3 years. Each participant completed three sessions spaced at least one week apart to assess intra-subject stability. During each session, they viewed 15 emotionally evocative movie clips, five each for positive, neutral, and negative emotions, and labeled their emotional responses accordingly. EEG signals were recorded using a 62-channel setup following the international 10--20 system, originally sampled at 1000~Hz and later downsampled to 128~Hz. Band Differential Entropy (BDE) features were extracted across five canonical frequency bands: delta, theta, alpha, beta, and gamma. A detailed electrode layout used in SEED is shown in Figure~\ref{fig:electrode_comparison}.(a).

\textbf{(2) DEAP:} The DEAP dataset~\cite{koelstra2012deap} comprises EEG and peripheral physiological recordings from 32 participants (17 male, 15 female) aged between 19 and 37 years. Each participant watched 40 one-minute music video clips and rated their emotional responses along four dimensions: Valence, Arousal, Dominance, and liking. EEG signals were recorded using a 32-channel BioSemi ActiveTwo system and downsampled to 128~Hz after preprocessing. The electrode layout used in DEAP is shown in Figure~\ref{fig:electrode_comparison}.(b).

\textbf{(3) DREAMER:} The DREAMER dataset~\cite{katsigiannis2018dreamer} comprises EEG and ECG recordings from 23 participants (14 male, 9 female) aged between 22 and 33 years. Participants watched 18 audio-visual movie clips and provided self-assessments of their emotional states using continuous ratings for Valence, Arousal, and Dominance. EEG signals were recorded using a 14-channel Emotiv EPOC device at a sampling rate of 128~Hz. Band Differential Entropy (BDE) features were extracted from overlapping time windows to capture temporal dynamics. The electrode layout used in DREAMER is shown in Figure~\ref{fig:electrode_comparison}.(c).

\subsection{Metrics}
To evaluate the performance of RBTransformer, we report the following classification metrics: Accuracy, Precision, Recall, and F1-score~\cite{powers2011evaluation}.

\textbf{(1) Accuracy:} Accuracy is the proportion of correct predictions across all samples, denoted in Eq.~\ref{eq:accuracy}:

\begin{equation} \label{eq:accuracy}
\mathit{Accuracy} = \frac{TP + TN}{TP + TN + FP + FN}
\end{equation}

Here, \(TP\), \(TN\), \(FP\), and \(FN\) represent true positives, true negatives, false positives, and false negatives, respectively.

\textbf{(2) Precision:} Precision~\cite{powers2011evaluation} is the proportion of true positives out of all predicted positives, denoted in Eq.~\ref{eq:precision}:

\begin{equation} \label{eq:precision}
\mathit{Precision} = \frac{TP}{TP + FP}
\end{equation}

\textbf{(3) Recall:} Recall~\cite{powers2011evaluation}, also known as Sensitivity or True Positive Rate, is the proportion of actual positives correctly identified by the model, denoted in Eq.~\ref{eq:recall}:

\begin{equation} \label{eq:recall}
\mathit{Recall} = \frac{TP}{TP + FN}
\end{equation}

\textbf{(4) F1-score:} The F1-score~\cite{powers2011evaluation} is the harmonic mean of Precision and Recall, balancing both in a single metric:

\begin{equation} \label{eq:f1}
\mathit{F1\text{-}score} = 2 \times \frac{\mathit{Precision} \times \mathit{Recall}}{\mathit{Precision} + \mathit{Recall}}
\end{equation}

\subsection{Training Configurations}
RBTransformer was trained using the PyTorch framework~\cite{paszke2019pytorch} on an NVIDIA P100 GPU. We used the AdamW optimizer~\cite{loshchilov2019decoupled} for a total of 300 epochs with an L2 regularization (weight decay) of $1\times10^{-3}$. A cosine annealing learning rate scheduler~\cite{loshchilov2017sgdr} dynamically adjusted the learning rate from $1\times10^{-3}$ to $1\times10^{-6}$ throughout the training. The model was trained using Cross-Entropy loss with label-smoothing~\cite{szegedy2016rethinking} set to 0.12. Batch size was set to 256 for the first 150 epochs and reduced to 64 for the remaining epochs. To handle class imbalance, SMOTE~\cite{chawla2002smote} was applied only to the training folds. A 5 fold cross-validation was carried out to evaluate the performance across all experiments which were tracked using Weights \& Biases (WandB)~\cite{wandb}.

\section{Results and Discussion} 
In this section, RBTransformer is extensively evaluated on the SEED, DEAP, and DREAMER datasets under a subject-dependent setting for both binary and multi-class classification tasks, and results are compared with existing state-of-the-art models. An extended metric evaluation is carried out for model performance validation, and to analyze class-wise predictions and visualize RBTransformer’s ability to distinguish and segregate class clusters in the latent space, confusion matrices and t-SNE plots are presented, respectively, across all datasets and dimensions for both classification settings. Finally, an ablation study is carried out to assess the impact of individual model components.

\subsection{Subject-Dependent Emotion Recognition Evaluation}
For performance evaluation, RBTransformer is assessed under a subject-dependent setting, in which the model is trained and evaluated separately on data from the same subjects. The entire dataset, which comprises EEG recordings across all subjects and trials is first preprocessed, and then is divided into training and validation sets using an 80–20 train–val split. A 5 fold cross-validation is then applied, where each fold maintains the subject-dependent setting. This means that within each fold, data from the same subjects can appear in both the training and validation splits, but always as distinct subsets. Same procedure is applied consistently across all three datasets, along all three dimensions, for both binary and multi-class classification tasks.

\subsection{Binary-Class Classification Evaluation Results}

RBTransformer is first evaluated under a binary-class classification setting across DEAP and DREAMER, along their three affective dimensions: Valence, Arousal, and Dominance. For binary classification, the labels are converted as follows: DEAP labels (1–9) are split into "High" (above 5) and "Low" (5 or below), while DREAMER labels (1–5) are split into "High" (above 3) and "Low" (3 or below). The performance of RBTransformer is compared against the existing state-of-the-art models, and the detailed results for both datasets under a binary-class classification setting are listed below.

% Match spacing with DREAMER table (narrower first column + tighter inter-column gap)
\begingroup
\setlength{\tabcolsep}{4pt} % default 6pt → 4pt for tighter spacing
\begin{table}[h]
    \centering
    \scriptsize % Compact font size
    \caption{Binary-Class Classification Performance Comparison on DEAP Dataset\label{tab:deap-binary}}
    \renewcommand{\arraystretch}{1.3} % Consistent row height
    \begin{adjustbox}{width=\columnwidth,center}
    \begin{tabularx}{\columnwidth}{>{\raggedright\arraybackslash}p{2.8cm} 
                                  >{\raggedright\arraybackslash}X 
                                  >{\raggedright\arraybackslash}X 
                                  >{\raggedright\arraybackslash}X}
        \toprule[1.2pt]
        \multicolumn{4}{c}{\textbf{DEAP Dataset}} \\
        \midrule[1.2pt]
        \textbf{Model} & \textbf{Valence} & \textbf{Arousal} & \textbf{Dominance} \\
        \midrule[1.2pt]
        ACRNN          & 93.72 ± 3.21 & 93.38 ± 3.73 & – \\
        GANSER         & 93.86 ± –    & 94.38 ± –    & – \\
        4D-CRNN        & 94.22 ± 2.61 & 94.58 ± 3.69 & – \\
        BiDCNN         & 94.38 ± 2.61 & 94.72 ± 2.56 & – \\
        CLDTA          & 94.58 ± 1.40 & 94.11 ± 2.10 & – \\
        DFCN           & 94.59 ± –    & 95.32 ± –    & 94.78 ± – \\
        RACNN          & 96.65 ± 2.65 & 97.11 ± 2.01 & – \\
        4D-ANN         & 96.90 ± 1.65 & 97.39 ± 1.75 & – \\
        TRPO-NET       & 97.87 ± 1.89 & 98.08 ± 1.83 & 98.33 ± 1.55 \\
        TDMNN          & 98.08 ± 2.13 & 98.25 ± 2.85 & – \\
        \textbf{RBTransformer (Ours)} & \textbf{99.84 ± 0.02} & \textbf{99.83 ± 0.05} & \textbf{99.82 ± 0.06} \\
        \bottomrule[1.2pt]
    \end{tabularx}
    \end{adjustbox}
\end{table}
\endgroup

% Reduce the inter-column spacing and make the first column a bit narrower
\begingroup
\setlength{\tabcolsep}{4pt} % default is 6 pt; tighten to 4 pt
\begin{table}[h]
    \centering
    \scriptsize % Compact font size
    \caption{Binary-Class Classification Performance Comparison on DREAMER Dataset\label{tab:dreamer-binary}}
    \renewcommand{\arraystretch}{1.3} % Consistent row height
    \begin{adjustbox}{width=\columnwidth,center}
    \begin{tabularx}{\columnwidth}{>{\raggedright\arraybackslash}p{2.8cm} % narrower Model col.
                                  >{\raggedright\arraybackslash}X 
                                  >{\raggedright\arraybackslash}X 
                                  >{\raggedright\arraybackslash}X}
        \toprule[1.2pt]
        \multicolumn{4}{c}{\textbf{DREAMER Dataset}} \\
        \midrule[1.2pt]
        \textbf{Model} & \textbf{Valence} & \textbf{Arousal} & \textbf{Dominance} \\
        \midrule[1.2pt]
        GANSER         & 85.28 ± –       & 84.16 ± –       & – \\
        DGCNN          & 86.23 ± 12.29   & 84.54 ± 10.18   & 85.02 ± 10.25 \\
        DFCN           & 93.15 ± –       & 91.30 ± –       & 92.04 ± – \\
        RACNN          & 95.55 ± 2.18    & 97.01 ± 2.74    & – \\
        ACRNN          & 97.93 ± 1.73    & 97.98 ± 1.92    & 98.23 ± 1.42 \\
        BiDCNN         & 98.35 ± 0.87    & 98.66 ± 1.46    & 99.01 ± 0.96 \\
        TRPO-NET       & 98.86 ± 0.57    & 98.97 ± 0.49    & 98.93 ± 0.69 \\
        TDMNN          & 99.45 ± 0.91    & 99.51 ± 0.79    & – \\
        \textbf{RBTransformer (Ours)} & \textbf{99.61 ± 0.05} & \textbf{99.74 ± 0.06} & \textbf{99.79 ± 0.04} \\
        \bottomrule[1.2pt]
    \end{tabularx}
    \end{adjustbox}
\end{table}
\endgroup

Experimental results under the binary-class classification setting are reported in Table~\ref{tab:deap-binary} and Table~\ref{tab:dreamer-binary} for the DEAP and DREAMER datasets, respectively, and are reported as mean accuracy ± standard deviation (ACC ± STD). On DEAP dataset, RBTransformer achieves ACC ± STD values of 99.84\% ± 0.02, 99.83\% ± 0.05, and 99.82\% ± 0.06 along it's Valence, Arousal, and Dominance dimensions, improving the previous state-of-the-art results by 1.76\% and 1.58\% on Valence and Arousal, respectively. Similarly, on DREAMER dataset, RBTransformer achieves ACC ± STD values of 99.61\% ± 0.05, 99.74\% ± 0.06, and 99.79\% ± 0.04 along it's Valence, Arousal, and Dominance dimensions, improving the previous state-of-the-art results by 0.16\% and 0.23\% on Valence and Arousal, respectively.

\subsection{Multi-Class Classification Evaluation Results}

Similarly, RBTransformer is also assessed on multi-class classification tasks across SEED, DEAP, and DREAMER. SEED is inherently a multi-class dataset with three emotion classes: Positive, Neutral, and Negative. For DEAP and DREAMER, the original continuous emotion ratings across three primary emotional dimensions, Valence, Arousal, and Dominance, are converted into discrete classes. Specifically, DEAP is converted into nine classes (1 to 9) for each dimension, while DREAMER is converted into five classes (1 to 5). RBTransformer's performance is directly compared against the existing state-of-the-art models on the task of multi-class classification, and the detailed results for each dataset are listed below.

% Group 1: SEED Dataset Table
\begingroup
\setlength{\tabcolsep}{3.5pt}
\renewcommand{\arraystretch}{1.25}

\begin{table}[h]
    \centering
    \scriptsize
    \caption{Multi-Class Classification Performance Comparison on SEED Dataset\label{tab:seed-multiclass}}
    \begin{adjustbox}{max width=\columnwidth,center}
    \begin{tabularx}{\linewidth}{@{\hspace{15pt}}%
                                 >{\raggedright\arraybackslash}p{2.8cm}%
                                 >{\raggedleft\arraybackslash}X%
                                 @{\hspace{20pt}}}
        \toprule[1.2pt]
        \multicolumn{2}{c}{\textbf{SEED Dataset}} \\
        \midrule[1.2pt]
        \textbf{Model} & \textbf{Accuracy} \\
        \midrule[1.2pt]
        DGCNN            & 90.40 ± 8.49 \\
        4D-CRNN          & 94.74 ± 2.32 \\
        CLDTA            & 95.09 ± 4.48 \\
        SST-EmotionNet   & 96.02 ± 2.17 \\
        4D-ANN           & 96.25 ± 1.86 \\
        TDMNN            & 97.20 ± 1.57 \\
        3DCANN           & 97.35 ± – \\
        GANSER           & 97.71 ± – \\
        \textbf{RBTransformer (Ours)} & \textbf{99.51 ± 0.02} \\
        \bottomrule[1.2pt]
    \end{tabularx}
    \end{adjustbox}
\end{table}
\endgroup

% Group 3: DEAP Dataset Table
\begingroup
\setlength{\tabcolsep}{4pt}
\renewcommand{\arraystretch}{1.3}

\begin{table}[h]
    \centering
    \scriptsize
    \caption{Multi-Class Classification Performance Comparison on DEAP Dataset\label{tab:deap-multiclass}}
    \begin{adjustbox}{width=\columnwidth,center}
    \begin{tabularx}{\columnwidth}{>{\raggedright\arraybackslash}p{2.8cm} 
                                  >{\raggedright\arraybackslash}X 
                                  >{\raggedright\arraybackslash}X 
                                  >{\raggedright\arraybackslash}X}
        \toprule[1.2pt]
        \multicolumn{4}{c}{\textbf{DEAP Dataset}} \\
        \midrule[1.2pt]
        \textbf{Model} & \textbf{Valence} & \textbf{Arousal} & \textbf{Dominance} \\
        \midrule[1.2pt]
        TRPO-NET             & 97.63 ± 2.38 & 97.74 ± 2.26 & 97.88 ± 2.24 \\
        \textbf{RBTransformer (Ours)} & \textbf{99.87 ± 0.04} & \textbf{99.84 ± 0.04} & \textbf{99.87 ± 0.05} \\
        \bottomrule[1.2pt]
    \end{tabularx}
    \end{adjustbox}
\end{table}
\endgroup

% Group 2: DREAMER Dataset Table
\begingroup
\setlength{\tabcolsep}{4pt}
\renewcommand{\arraystretch}{1.3}

\begin{table}[h]
    \centering
    \scriptsize
    \caption{Multi-Class Classification Performance Comparison on DREAMER Dataset\label{tab:dreamer-multiclass}}
    \begin{adjustbox}{width=\columnwidth,center}
    \begin{tabularx}{\columnwidth}{>{\raggedright\arraybackslash}p{2.8cm} 
                                  >{\raggedright\arraybackslash}X 
                                  >{\raggedright\arraybackslash}X 
                                  >{\raggedright\arraybackslash}X}
        \toprule[1.2pt]
        \multicolumn{4}{c}{\textbf{DREAMER Dataset}} \\
        \midrule[1.2pt]
        \textbf{Model} & \textbf{Valence} & \textbf{Arousal} & \textbf{Dominance} \\
        \midrule[1.2pt]
        TRPO-NET             & 98.18 ± 0.97 & 98.37 ± 0.93 & 98.40 ± 0.80 \\
        \textbf{RBTransformer (Ours)} & \textbf{99.54 ± 0.06} & \textbf{99.55 ± 0.04} & \textbf{99.60 ± 0.05} \\
        \bottomrule[1.2pt]
    \end{tabularx}
    \end{adjustbox}
\end{table}
\endgroup

Experimental results under the multi-class classification setting are reported in Table~\ref{tab:seed-multiclass}, Table~\ref{tab:deap-multiclass}, and Table~\ref{tab:dreamer-multiclass} for the SEED, DEAP, and DREAMER datasets, respectively, and are reported as mean accuracy ± standard deviation (ACC ± STD). On SEED dataset, RBTransformer achieves an ACC ± STD of 99.51\% ± 0.02, improving the previous state-of-the-art results by 1.80\%. On DEAP dataset, it achieves ACC ± STD values of 99.87\% ± 0.04, 99.84\% ± 0.04, and 99.87\% ± 0.05 along it's Valence, Arousal, and Dominance dimensions, respectively, improving the previous state-of-the-art results by 2.10\%, 2.24\%, and 1.99\%. Similarly, on DREAMER dataset, RBTransformer achieves ACC ± STD values of 99.54\% ± 0.06, 99.55\% ± 0.04, and 99.60\% ± 0.05 along it's Valence, Arousal, and Dominance dimensions, improving the previous state-of-the-art results by 1.18\%, 1.36\%, and 1.20\%, respectively.

% \begin{figure}[htbp]
% \centering
% \includegraphics[width=\linewidth]{images/ablations.png}
% \caption{Ablation study for binary classification on the DREAMER dataset (Arousal dimension). Top: Impact of inter-cortical attention. Bottom: Impact of training and regularization components including ADASYN, SMOTE with label smoothing, weight decay, and dropout.}
% \label{fig:ablation_results}
% \end{figure}

\subsection{Ablation Study}
To evaluate the contribution of different architectural components, training configurations, and regularization choices for RBTransformer, five ablation experiments are carried out on the DREAMER dataset along the Arousal dimension for the binary classification task. Results are reported as the mean and standard deviation over five cross-validation folds. In the first ablation, the Inter-Cortical Attention mechanism is removed from the model to assess its impact. 

\begin{figure}[htbp]
\centering
\includegraphics[width=\linewidth]{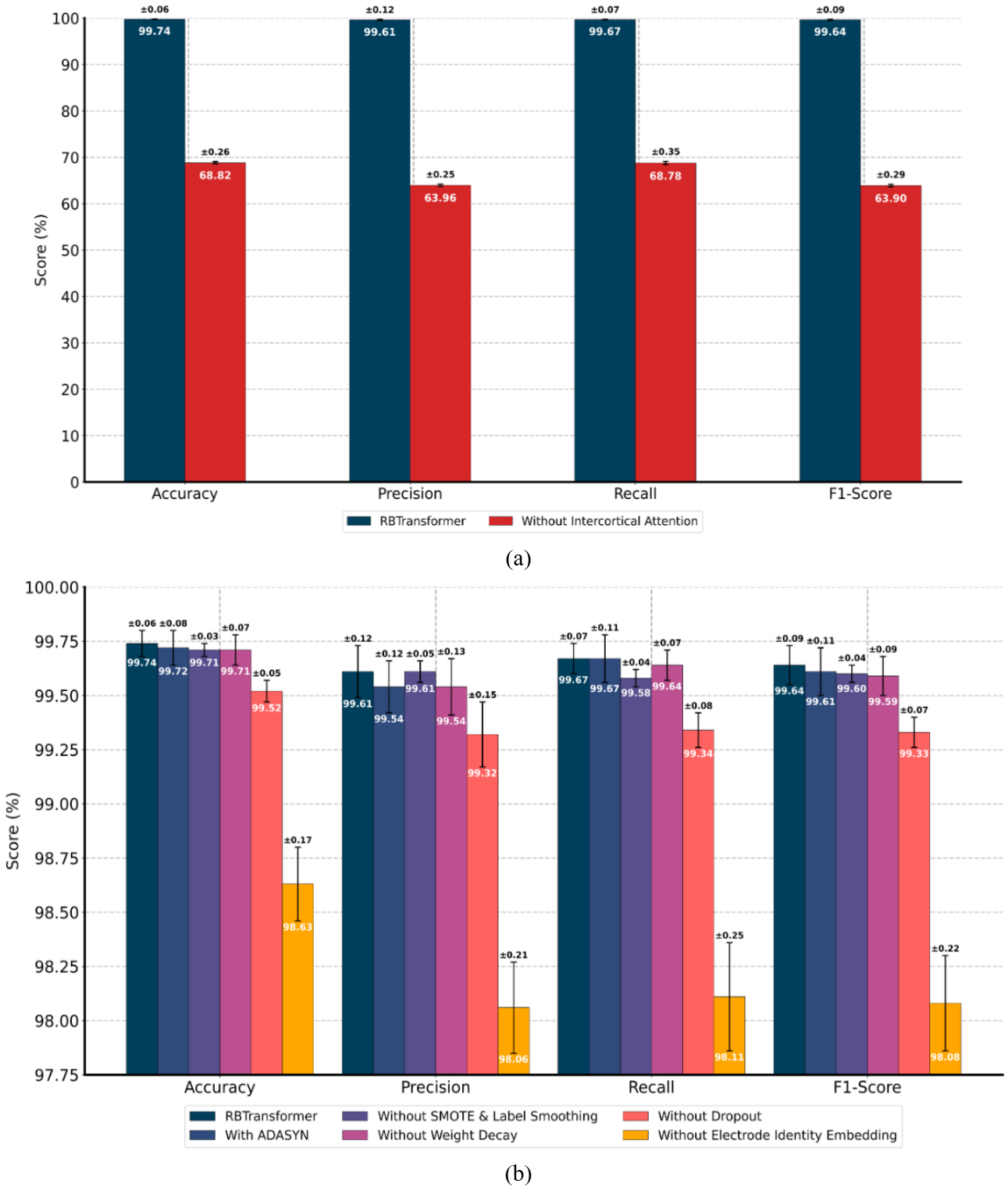}
\caption{Ablation study for binary classification on the DREAMER dataset (Arousal dimension). Top: Impact of inter-cortical attention. Bottom: Impact of training and regularization components including ADASYN, SMOTE with label smoothing, weight decay, and dropout.}
\label{fig:ablation_results}
\end{figure}

As shown in Figure~\ref{fig:ablation_results}.(a), this leads to a sharp drop of around 30\% across all four metrics, highlighting the critical importance of the Inter-Cortical Attention mechanism. Figure~\ref{fig:ablation_results}.(b) presents the remaining four ablations: the second ablation replaces SMOTE with ADASYN; the third disables both SMOTE and label smoothing; the fourth removes weight decay; and the fifth removes dropout. In all cases, the performance ranks lower than the final RBTransformer configuration, demonstrating the effectiveness and necessity of the chosen architectural and training design decisions.

\begin{table*}[!t]
\caption{Extended Metric Evaluation (Accuracy, Precision, Recall, F1-Score), Mean ± SD on SEED, DEAP, and DREAMER Datasets (Binary and Multi-Class Classification)}
\label{tab:extended_metric_evaluation}
\centering
\renewcommand{\arraystretch}{1.3} % Increase row padding for clarity
\setlength{\tabcolsep}{8pt} % Adjust column padding
\resizebox{\textwidth}{!}{
\begin{tabular}{l|c|ccc|ccc|ccc|ccc}
\hline
\multirow{3}{*}{\textbf{Metric}} & \multirow{3}{*}{\textbf{SEED}} & \multicolumn{6}{c|}{\textbf{DEAP}} & \multicolumn{6}{c}{\textbf{DREAMER}} \\
\cline{3-14}
 & & \multicolumn{3}{c|}{\makecell{\textbf{Binary-Class} \\ \textbf{Classification}}} & 
     \multicolumn{3}{c|}{\makecell{\textbf{Multi-Class} \\ \textbf{Classification}}} & 
     \multicolumn{3}{c|}{\makecell{\textbf{Binary-Class} \\ \textbf{Classification}}} & 
     \multicolumn{3}{c}{\makecell{\textbf{Multi-Class} \\ \textbf{Classification}}} \\
\cline{3-14}
 & & \textbf{Valence} & \textbf{Arousal} & \textbf{Dominance} & \textbf{Valence} & \textbf{Arousal} & \textbf{Dominance} & \textbf{Valence} & \textbf{Arousal} & \textbf{Dominance} & \textbf{Valence} & \textbf{Arousal} & \textbf{Dominance} \\
\hline
Accuracy (\%)   & 99.51 ± 0.02 & 99.84 ± 0.02 & 99.83 ± 0.05 & 99.82 ± 0.06 & 99.87 ± 0.04 & 99.84 ± 0.04 & 99.87 ± 0.05 & 99.61 ± 0.05 & 99.74 ± 0.06 & 99.79 ± 0.04 & 99.54 ± 0.06 & 99.55 ± 0.04 & 99.60 ± 0.05 \\
Precision (\%)  & 99.51 ± 0.02 & 99.84 ± 0.02 & 99.83 ± 0.06 & 99.81 ± 0.07 & 99.87 ± 0.06 & 99.84 ± 0.05 & 99.86 ± 0.06 & 99.58 ± 0.05 & 99.61 ± 0.12 & 99.63 ± 0.06 & 99.54 ± 0.06 & 99.49 ± 0.06 & 99.56 ± 0.05 \\
Recall (\%)     & 99.51 ± 0.02 & 99.83 ± 0.02 & 99.83 ± 0.06 & 99.81 ± 0.06 & 99.86 ± 0.03 & 99.84 ± 0.05 & 99.87 ± 0.04 & 99.60 ± 0.06 & 99.67 ± 0.07 & 99.72 ± 0.09 & 99.54 ± 0.05 & 99.46 ± 0.09 & 99.58 ± 0.14 \\
F1-score (\%)   & 99.51 ± 0.02 & 99.83 ± 0.02 & 99.83 ± 0.06 & 99.81 ± 0.07 & 99.86 ± 0.04 & 99.84 ± 0.04 & 99.87 ± 0.05 & 99.59 ± 0.06 & 99.64 ± 0.09 & 99.68 ± 0.06 & 99.54 ± 0.05 & 99.47 ± 0.07 & 99.57 ± 0.09 \\
\hline
\end{tabular}
}
\end{table*}

% Forcing the figure to the top of the section without going to the next page
\begin{figure*}[!t]
    \centering
    \includegraphics[width=\textwidth]{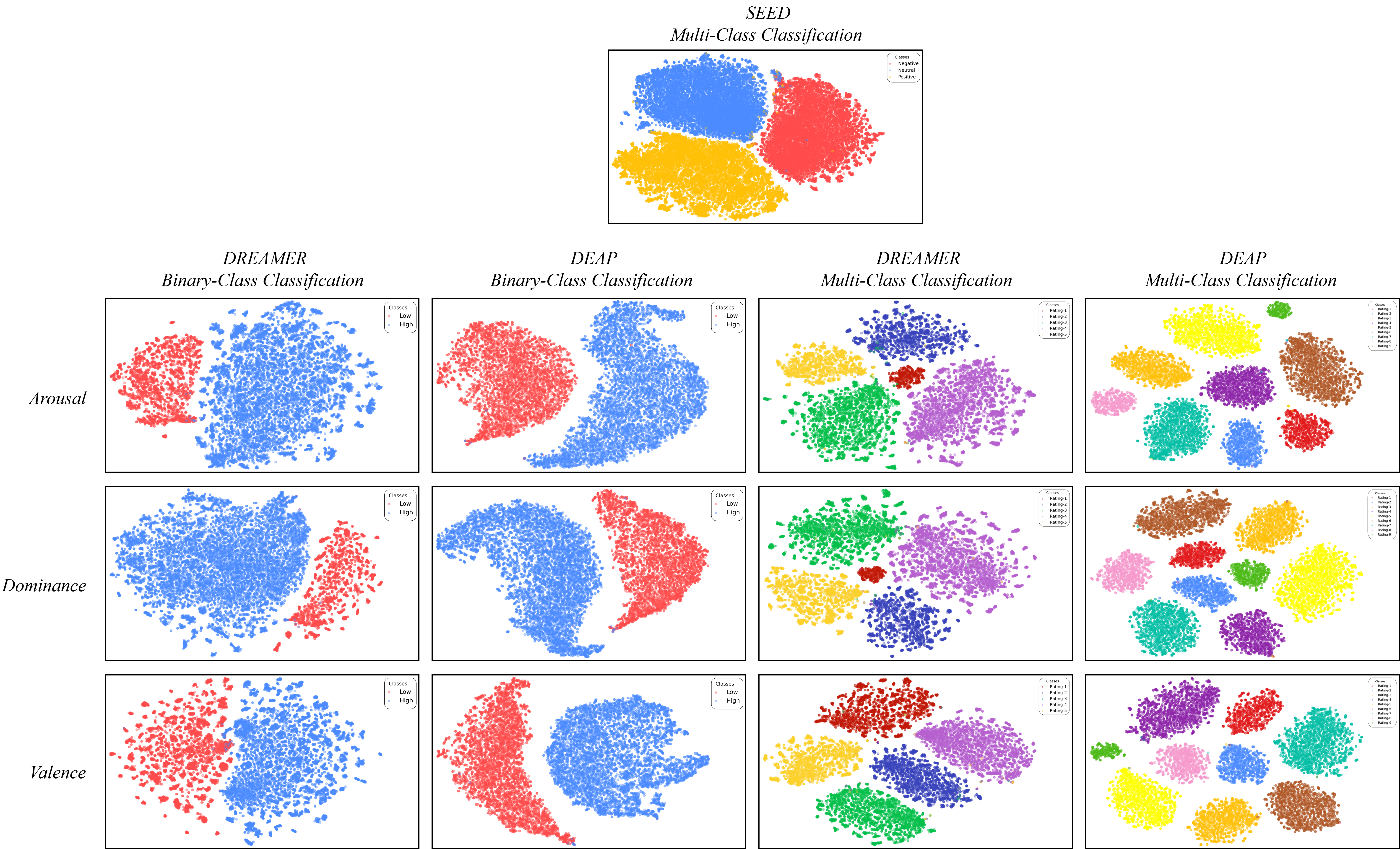}
    \caption{t-SNE visualization of RBTransformer's learned feature representations on SEED, DEAP, and DREAMER for both binary and multi-class classification tasks across all dimensions (Valence, Arousal, and Dominance).}
    \label{fig:tsne-plots}
\end{figure*}

\subsection{Extended Metric Evaluation for Performance Validation}
While accuracy provides a general indication of performance, it may overlook model's class-wise prediction capability. So, to assess the robustness and generalizability of RBTransformer in depth, an extended evaluation of metrics, including Precision, Recall, and F1-Score is carried out. Precision reflects the model’s ability to avoid false positives, Recall captures how well it retrieves relevant instances, and F1-Score balances the two. Evaluating these metrics helps assess whether the model performs reliably across all emotional classes, including less frequent ones. The extended metric evaluation is carried out on the SEED, DEAP, and DREAMER datasets across the three affective dimensions, Valence, Arousal, and Dominance (for DEAP and DREAMER), for both binary and multi-class classification tasks. Results are reported as the mean and standard deviation over five cross-validation folds and include Accuracy, Precision, Recall, and F1-Score. As shown in Table 1, RBTransformer demonstrates balanced, stable, and consistently high performance across all metrics, datasets, and corresponding dimensions, supporting its overall accuracy and ensuring that the observed accuracy is not driven by biased learning or class imbalance.

\begin{figure*}[!t]
    \centering
    \includegraphics[width=0.94\textwidth]{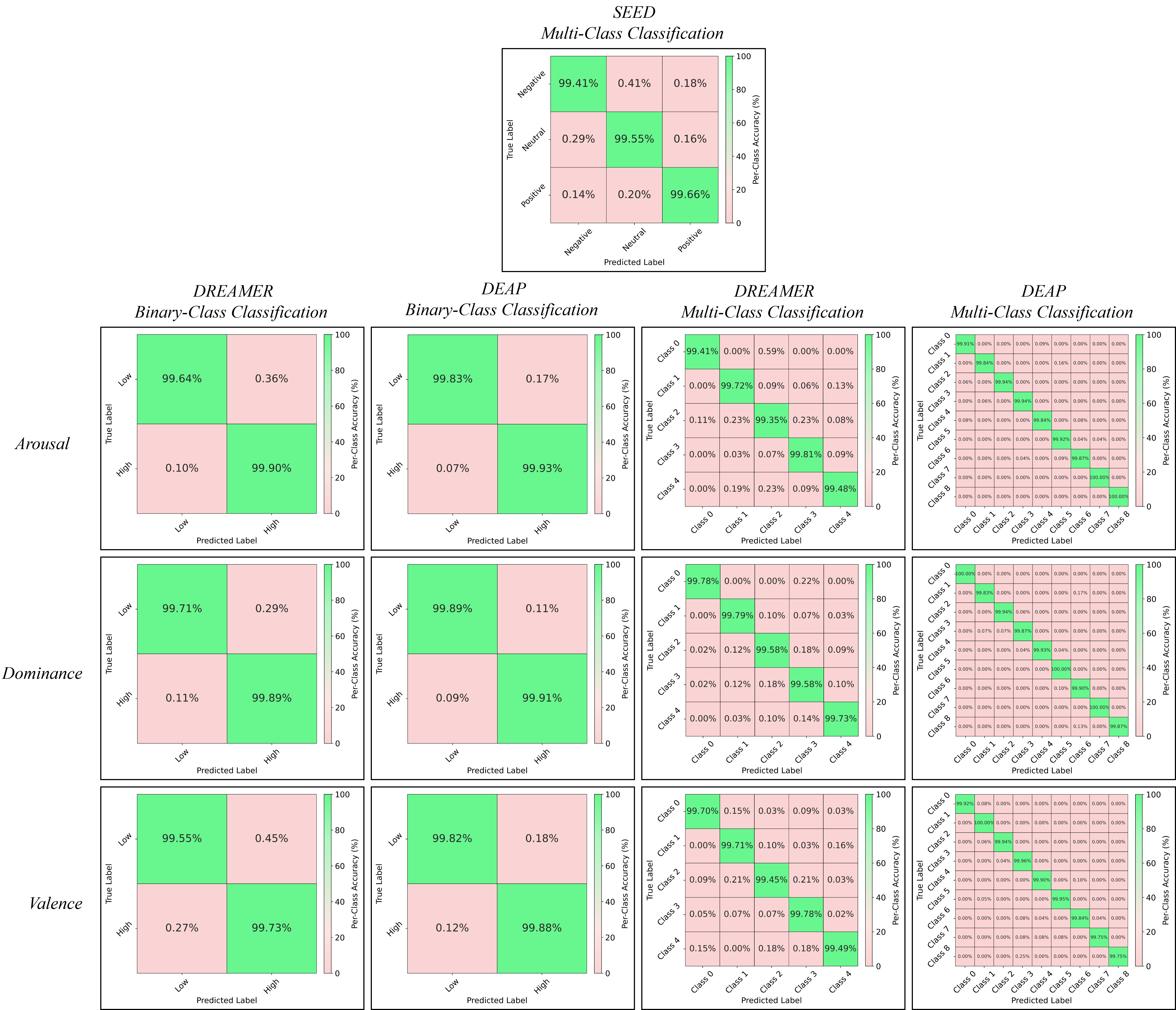}
    \caption{Confusion matrices of RBTransformer predictions on SEED, DEAP, and DREAMER for both binary and multi-class classification across all dimensions (Valence, Arousal, and Dominance).}
    
    \label{fig:confusion_matrices}
\end{figure*}
\subsection{Visualization of Learned Feature Representations by RBTransformer}
\label{sec:tsne-visualization}
To visualize RBTransformer’s ability to classify unseen EEG signals, its high-dimensional operational feature space is projected into two dimensions using the t-SNE (t-Distributed Stochastic Neighbor Embedding) algorithm. As observable from Figure~\ref{fig:tsne-plots}, RBTransformer demonstrates strong performance across all datasets: SEED shows clear separation of Positive, Neutral, and Negative classes; in binary classification, DEAP and DREAMER form distinct groupings for High and Low classes across all dimensions, Valence, Arousal, and Dominance; and in multi-class settings, DEAP’s nine and DREAMER’s five emotion classes also form well-segregated clusters across these same dimensions, validating the model’s ability to capture and differentiate emotional states in the learned feature space.

% % Forcing the figure to the top of the section without going to the next page
% \begin{figure*}[!t]
%     \centering
%     \includegraphics[width=\textwidth]{images/tsne-plots-rbtransformer.png}
%     \caption{t-SNE visualization of RBTransformer's learned feature representations on SEED, DEAP, and DREAMER for both binary and multi-class classification tasks across all dimensions (Valence, Arousal, and Dominance).}
%     \label{fig:tsne-plots}
% \end{figure*}

\subsection{Confusion Matrix Analysis}
In addition to plotting t-SNE plots for visualizing RBTransformer’s ability to distinguish between different class clusters in the latent space, we plot confusion matrices for all the datasets to further analyze the class-wise prediction capability of RBTransformer in detail. The confusion matrices are generated for all three datasets, SEED, DEAP, and DREAMER, across all three affective dimensions, Valence, Arousal, and Dominance (for DEAP and DREAMER), under both binary and multi-class classification tasks, shown in Figure~\ref{fig:confusion_matrices}. As observable from the plots, RBTransformer achieves a very high alignment between predicted and true labels, with most values concentrated along the principal diagonal. On average, the model achieves class-wise accuracy of around 99.50\%, with only minor misclassifications appearing in off-principal-diagonal entries, providing a quantitative and visual confirmation of RBTransformer’s strong class-wise performance across all datasets and affective dimensions under both classification settings.

% Conclusion
\section{Conclusion}
In this paper, we propose RBTransformer, a Transformer-based neural network architecture that simulates and captures inter-cortical neural interactions in latent space for effective EEG-based emotion recognition. RBTransformer first converts raw EEG signals into Band Differential Entropy (BDE) tokens that are then passed into an Electrode-Identity Embedding, which allows the model to retain spatial awareness of each electrode’s position and ordering. These intermediate features are then passed through an Inter-Cortical Attention module, which is made up of successive Inter-Cortical Multi-Head Attention blocks, each comprised of electrode × electrode attention matrix, allowing each electrode to interact directly with every other electrode in the latent space, and a feedforward network. These attention blocks are stacked on top of one another, which mimics the recurrent exchange of signals between distinct brain regions, allowing RBTransformer to capture both inter-regional spatial dependencies and localized temporal dependencies, without relying on any handcrafted features or explicit temporal sequence modeling. Experimental results demonstrate that RBTransformer achieves state-of-the-art performance, outperforming all existing models, across all benchmarks, DEAP, DREAMER, and SEED, across all dimensions, Valence, Arousal, and Dominance (for DEAP and DREAMER), under both binary and multi-class EEG-based emotion recognition tasks.

% References
\bibliographystyle{IEEEtran}
\bibliography{references}

\end{document}